\documentclass{PoS}

\title{Infrared behavior and infinite-volume limit of gluon and ghost 
       propagators in Yang-Mills theories}

\ShortTitle{Infinite-volume limit of infrared gluon and ghost propagators}

\author{Attilio Cucchieri\\
  E-mail: \email{attilio@ifsc.usp.br}}

\author{\speaker{Tereza Mendes}\\
  E-mail: \email{mendes@ifsc.usp.br}\\\\
  Instituto de F\'\i sica de S\~ao Carlos, Universidade de S\~ao Paulo,\\
  Caixa Postal 369, 13560-970 S\~ao Carlos, SP, Brazil}

\abstract{
Lattice studies of the infrared behavior of gluon and ghost
propagators are a key probe of confinement
scenarios in Yang-Mills theories. However, finite-volume
effects become an important issue as the infrared limit is
approached. By considering general quantities --- namely an 
associated susceptibility in the gluon case and properties of 
the lowest-lying eigenmode of the Faddeev-Popov matrix in the 
ghost case --- one can derive rigorous upper and lower 
bounds for the propagators. The bounds allow a better 
control over the extrapolation of lattice results to the infinite-volume 
limit. In the case of the gluon propagator, an intuitive 
statistical argument suggests a precise volume dependence for
the bounds. This dependence is nicely confirmed by the lattice 
data, leading to a finite gluon propagator at zero momentum.
At the same time, an enhancement of the ghost propagator in the infrared 
limit seems unlikely. Our analysis is applied to the case of Landau gauge 
and SU(2) gauge group, using the largest lattice sizes to date.
}

\FullConference{8th Conference Quark Confinement and the Hadron Spectrum \\
		 September 1-6 2008\\
		 Mainz, Germany}

\begin{document}

\section{Introduction}

Confinement in QCD and in Yang-Mills theories in general
is associated with long-range (infrared) effects. It is thus
necessary to study the infrared behavior of the theory's Green's 
functions using nonperturbative methods. These studies include 
numerical (lattice) as well as analytic methods and consider basic 
(gauge-dependent) quantities --- such as gluon and ghost propagators 
--- in order to test the predictions of the
so-called confinement scenarios (see e.g.\ \cite{Maas:2006qw}
and references therein). In the case of lattice simulations, one
has at one's disposal a true first-principles method, with no uncontrolled
approximations. On the other hand, extreme care must be taken to extract
the true infrared behavior of the propagators from lattice data, since
significant systematic errors may affect the extrapolations that are 
needed in order to get physical results. The most important such errors 
are Gribov-copy effects\footnote{The problem of Gribov-copy effects has 
been extensively studied on the lattice 
\cite{Cucchieri:1997dx,Cucchieri:1997ns,Silva:2004bv,Maas:2008ri}.
We comment briefly on this issue in our Conclusions.}
 --- related to the fact that the relevant objects 
are gauge-dependent quantities --- and finite-size effects. The latter
is especially important in the investigation of the infrared limit, since
the smallest nonzero momentum that can be represented on a lattice of
linear extension $L$ is proportional to $1/L$. Thus, a sensible range of
small momenta can only be properly simulated on a very large lattice.
Here, we consider carefully the elimination of finite-size effects 
through a better control of the extrapolation of our data to the
infinite-volume limit, as described below. 

The extrapolation of gluon- and ghost-propagator data to
infinite lattice volume is a delicate task, since the correct volume
dependence of the data may not be easily inferred from the behavior
on medium-size lattices and since some quantities, such as the zero-momentum 
gluon propagator, are quite noisy. For these reasons it proves very helpful 
to obtain constraints on the infrared behavior of the propagators, 
as the upper and lower bounds presented here.
We remark that these bounds are valid at each lattice volume $V$ and
must be extrapolated to infinite volume, just as for the propagators.
The additional advantage, besides establishing a range of allowed values 
for the propagators, is that the bounds are written in terms of ``friendly'' 
quantities --- i.e.\ easier to compute, better behaved or more intuitive than 
the propagators themselves. It will therefore be more convenient 
to study the volume dependence of the bounds first, in order to assess 
the volume dependence of the propagators.
We describe and apply the gluon and ghost bounds --- in pure $SU(2)$
theory and Landau gauge --- respectively
in Sections \ref{gluon} and \ref{ghost} below. As can be seen
from our analysis, we obtain a finite nonzero gluon propagator 
and a tree-level-like ghost propagator in the infrared limit. Possible 
implications of these results for the currently accepted confinement 
scenarios are discussed in the Conclusions.

\section{Gluon Bounds}
\label{gluon}

Rigorous upper and lower bounds have been introduced in Ref.\ 
\cite{Cucchieri:2007rg} for the gluon propagator at zero momentum, 
defined as
\begin{equation}
D(0) \; = \; \frac{V}{d (N_c^2 - 1)} \sum_{\mu, b} 
\langle | {\widetilde A}^b_{\mu}(0) |^2 \rangle \; ,
\end{equation}
where $ {\widetilde A}^b_{\mu}(p) $ is the Fourier transform of the
gluon field $A^b_{\mu}(x)$ in pure $SU(N_c)$ gauge theory, 
$\langle \cdot \rangle$ stands for the 
path integral (Monte Carlo) average, $V = N^d$ is the lattice volume
and we consider $d$ space-time dimensions. Let us define the quantity
\begin{equation}
{M}(0) \, = \, \frac{1}{d (N_c^2 - 1)}
   \sum_{b,\mu} | {\widetilde A}^b_{\mu}(0) | \; .
\label{eq:mag}
\end{equation}
It is straightforward to show that this quantity is related 
to $D(0)$ as
\begin{equation}
V \, {\langle {M}(0) \rangle}^2 \, \leq \; D(0)\;
    \leq \; V d (N_c^2 - 1) \, \langle {{M}(0)}^2 \rangle \; ,
\label{eq:Dbounds}
\end{equation}
which provides us with rigorous upper and lower bounds for $D(0)$ that 
must be satisfied at every volume $V$.
We can now try to interpret the quantities ${\langle {M}(0) \rangle}^2$ and
$\langle {{M}(0)}^2 \rangle$, to obtain perhaps an understanding of their 
volume dependence. We start by noting that if we take the above
``magnetization'' without the absolute value, i.e.\ considering
\begin{equation}
{M}'(0) \, = \, \frac{1}{d (N_c^2 - 1)}
   \sum_{b,\mu} {\widetilde A}^b_{\mu}(0) \; ,
\end{equation}
we get a null Monte Carlo average: ${\langle {M}'(0) \rangle} = 0$.
Because of the absolute value, the quantity defined in Eq.\ (\ref{eq:mag}) 
has a nonzero average at finite $V$, but it should go to zero 
at least as fast as $V^{-1/d}$, as shown in \cite{Zwanziger:1990by}.
We now note that 
$\, V \langle {{M}(0)}^2 \rangle $ 
is essentially the susceptibility associated with the magnetization ${M}'(0)$ 
(since the average of this magnetization is zero). 
For a $d$-dimensional spin system one thus
expects to see $ \,V \langle {{M}(0)}^2  \rangle \sim const$, i.e.\ the 
statistical variance of the magnetization is proportional to the inverse
of the volume, a behavior known as {\em self-averaging}.
At the same time, considering the statistical fluctuations in the Monte Carlo 
sampling of ${M}(0)$, we would expect ${\langle {M}(0) \rangle}^2 $ to have 
the same volume dependence
as $\langle {{M}(0)}^2 \rangle$ \cite{Cucchieri:2007rg}. 

The simple statistical argument presented above suggests that both
${\langle {M}(0) \rangle}^2$ and $\langle {{M}(0)}^2 \rangle$ should show a 
volume dependence as $1/V$, implying (for $d>2$) a much stronger approach
to zero than the limiting behavior for $M(0)$ obtained in \cite{Zwanziger:1990by}
and mentioned earlier. On the other hand, the suppression with $1/V$ is 
compensated by the volume factor for both bounds in Eq.\ (\ref{eq:Dbounds}).
Consequently, if this suggested behavior for the susceptibilities is verified, 
$D(0)$ converges to a nonzero constant in the infinite-volume limit.
Note that the bounds in Eq.\ (\ref{eq:Dbounds}) apply to any gauge and 
that they can be immediately extended to the case $D(p)$ with $p \neq 0$.

We have investigated the volume dependence of the bounds for pure $SU(2)$ 
gauge theory in Landau gauge, considering physical lattice volumes of up to
$\,a^4 V \approx (27 \,\mbox{fm})^4$. We find remarkably good agreement with
the predicted $1/V$ behavior for 
${\langle {M}(0) \rangle}^2$ and $\langle {{M}(0)}^2 \rangle$, as can be seen
in plots and tables in Ref.\ \cite{Cucchieri:2007rg}. More precisely, by 
fitting the two quantities to $1/V^{\alpha}$ we get the
exponents $\alpha$ respectively 0.995(10) and 0.998(10). 
Analogously, an analysis for the $SU(3)$ case (considering somewhat smaller
volumes) yields the exponents 1.058(6) and 1.056(6) \cite{Oliveira:2008uf}.
A similar behavior is also obtained by a study introducing a modified 
gauge-fixing procedure (in order to check for possible Gribov-copy effects) 
\cite{flip}.
Finally, a finite nonzero gluon propagator has been recently
obtained using improved actions and anisotropic lattices \cite{Gong:2008td}.
We remark that this behavior has also been clearly observed on very large lattices
in $3d$ \cite{Cucchieri:2003di,Cucchieri:2007rg}
but not in $2d$ \cite{Maas:2007uv,Cucchieri:2007rg}.

\section{Ghost Bounds}
\label{ghost} 

Rigorous lower and upper bounds for the ghost propagator $G(p)$
were proposed in \cite{Cucchieri:2008fc}. We recall that $G(p)$ is
given by the inverse of the Faddeev-Popov (FP) matrix ${\cal M}$
and that an infrared enhancement of $G(p)$ with respect to the tree-level 
ghost propagator $G(p)\sim p^{-2}$ is generally expected as a sign of
confinement. By straightforward calculations --- independent of the ones
performed in the gluon case --- we can establish bounds for the ghost 
propagator. In Landau gauge, for any nonzero momentum $p$, one finds
\begin{equation}
\frac{1}{N_c^2 - 1} \,  \frac{1}{\lambda_{min}} \, \sum_a \,
  | {\widetilde \psi_{min}(a,p)} |^2 \,
              \leq \, G(p) \, \leq \, \frac{1}{\lambda_{min}} \; ,
\label{eq:Gineq}
\end{equation}
where $\lambda_{min}$ is the smallest nonzero eigenvalue of the FP
operator ${\cal M}$ and ${\widetilde \psi_{min}(a,p)}$ is the corresponding 
eigenvector. Note that the upper bound is independent of the momentum $p$.
If we now assume $\lambda_{min}\sim L^{-\nu}$ and
$\,G(p) \sim p^{-2-2\kappa}$ at small $p$, we have that  
$2+2\kappa \leq \nu$, i.e.\ $\nu > 2$, is a necessary condition for
the infrared enhancement of $G(p)$. A similar analysis can be carried 
out \cite{Cucchieri:2006hi} for a generic gauge condition.
Consider the Gribov region $\Omega$, where all eigenvalues
of ${\cal M}$ are positive. In the infinite-volume limit,
entropy favors configurations near the Gribov horizon $\partial \Omega$
(where $\lambda_{min}$ goes to zero). Thus, inequalities such as (\ref{eq:Gineq})
can tell us if one should expect an enhancement $G(p)$
when the Boltzmann weight gets concentrated\footnote{For example, 
in $4d$ Maximally Abelian gauge one sees
that $\lambda_{min}$ goes to zero at large volume but the ghost propagator
stays finite at zero momentum \cite{MAG}.} on $\partial \Omega \,$.

Our study in the $SU(2)$ Landau case \cite{Cucchieri:2008fc}, using the very large 
lattices mentioned in the previous section, suggests that $\nu<2$ (for $d=4$). 
This tree-level-like behavior is confirmed if one considers the dressing function 
$p^2 G(p)$. Indeed, the data for the dressing function
can be well fitted by $a - b \log(1 + c p^2)$ \cite{Cucchieri:2008fc}, 
supporting $\kappa = 0$.
This is also observed in $d=3$. For $d=2$ enhancement is observed, with
a behavior $\sim p^{-2 \kappa}$ and $\kappa$ between 0.1 and 0.2
\cite{Maas:2007uv,Cucchieri:2008fc}.

\section{Conclusions}

By using rigorous upper and lower bounds to
constrain the infrared behavior of gluon and ghost propagators,
we obtain a finite nonzero gluon propagator at zero momentum and
an essentially constant ghost dressing function in the infrared
limit. These results seem to contradict the commonly
accepted confinement scenarios of Gribov-Zwanziger and Kugo-Ojima
\cite{Cucchieri:2006xi}. However, as pointed out in \cite{Cucchieri:2008yp},
the above results are not completely in disagreement with the 
Gribov-Zwanziger approach.
In particular, it has been recently shown \cite{3d4d} that using 
the Gribov-Zwanziger approach, i.e.\ by restricting the functional integration 
to the Gribov region $\Omega$, one can also obtain in $3d$ and
$4d$ a finite nonzero gluon propagator and a tree-level-like ghost propagator 
in the infrared limit. It is interesting that the same approach cannot be 
applied to the $2d$ case \cite{Dudal:2008xd}.
Let us also note \cite{Cucchieri:2008yp} that even though the
Gribov-Zwanziger and the Kugo-Ojima confinement scenarios seem to predict 
similar infrared behavior for the propagators, it is not clear how to relate the 
(Euclidean) cutoff at the Gribov horizon to the (Minkowskian) approach of 
Kugo-Ojima \cite{Zwanziger:2003cf}.

Similar results for the gluon and ghost propagators are
obtained by various groups using very large lattice 
volumes \cite{largevolume}, both in the $SU(2)$ and in the $SU(3)$ cases. 
[The equivalence between the infrared propagators in $SU(2)$ and $SU(3)$ 
gauge theories can be seen e.g.\ in \cite{Cucchieri:2007zm}.]
Of course, one should also recall that the region $\Omega$ is actually not 
free of Gribov copies and that the configuration space should be identified 
with the so-called fundamental modular region $\Lambda$. On the other hand, 
the restriction to $\Lambda$ and the numerical 
verification of the Gribov-Zwanziger scenario are separate issues
\cite{Cucchieri:2008yp}. 
Indeed, this scenario is based on the restriction of
the configuration space to the region $\Omega$, which includes $\Lambda$. 
Finally, as explained in \cite{Cucchieri:1997dx}, the restriction to 
$\Lambda$ can only make the ghost propagator less singular, as confirmed by 
recent lattice data \cite{Maas:2008ri}.

\noindent{\bf Acknowledgements}
The authors acknowledge partial support from the Brazilian Funding
Agencies FAPESP and CNPq. T.M. also thanks the Theory Group at DESY-Zeuthen
for hospitality and the Alexander von Humboldt Foundation for financial
support.

\end{document}